\documentclass[acmsmall]{acmart}

\usepackage{csquotes}
\usepackage{subcaption}
\usepackage{todonotes}
\usepackage{longtable}
\usepackage{multirow}
\usepackage{array}
\newcolumntype{R}[1]{>{\raggedleft\let\newline\\\arraybackslash\hspace{0pt}}m{#1}}
\usepackage[flushleft]{threeparttable}

\makeatletter
\patchcmd{\csq@bquote@i}{{#6}}{{\emph{#6}}}{}{}
\makeatother

\AtBeginDocument{%
  }

\begin{document}



\title{Secret Use of Large Language Model (LLM)}



\author{Zhiping Zhang}
\email{zhang.zhip@northeastern.edu}
\affiliation{%
  \institution{Northeastern University}
  \city{Boston}
  \state{MA}
  \country{USA}
}

\author{Chenxinran Shen}
\email{elise.shen007@gmail.com}
\affiliation{%
  \institution{Northeastern University}
   \city{Boston}
   \state{MA}
  \country{USA}
}

\author{Bingsheng Yao}
\email{b.yao@northeastern.edu}
\affiliation{%
  \institution{Northeastern University}
  \city{Boston}
  \state{MA}
  \country{USA}
}

\author{Dakuo Wang}
\email{d.wang@neu.edu}
\affiliation{%
    \institution{Northeastern University}
    \city{Boston}
    \state{MA}
    \country{USA}
}

\author{Tianshi Li}
\email{tia.li@northeastern.edu}
\orcid{0000-0003-0877-5727}
\affiliation{%
    \institution{Northeastern University}
    \city{Boston}
    \state{MA}
    \country{USA}
}

\renewcommand{\shortauthors}{Zhang et al.}
\renewcommand{\shorttitle}{}
\newcommand{\surveyallusernumber}{180}
\newcommand{\surveynoncernedusernumber}{68}
\newcommand{\surveyconcernedusernumber}{112}
\newcommand{\surveycasenumber}{125}
\newcommand{\experimentrawdatapoint}{318}
\newcommand{\experimentvalidsamplesize}{300}
\newcommand{\tl}[1]{\textcolor{blue}{\bf [*** TL: #1]}}
\newcommand{\revision}[1]{#1}
\renewcommand{\sectionautorefname}{Section}
\renewcommand{\subsectionautorefname}{Section}
\renewcommand{\subsubsectionautorefname}{Section}

\begin{abstract}
The advancements of Large Language Models (LLMs) have decentralized the responsibility for the transparency of AI usage.
Specifically, LLM users are now encouraged or required to disclose the use of LLM-generated content for varied types of real-world tasks.
However, an emerging phenomenon, users' \textbf{secret use of LLM}, raises challenges in ensuring end users adhere to the transparency requirement.
Our study used mixed-methods with an exploratory survey (\surveycasenumber{} real-world secret use cases reported)
and a controlled experiment among \experimentvalidsamplesize{} users to investigate the contexts and causes behind the secret use of LLMs.
We found that such secretive behavior is often triggered by certain tasks, transcending demographic and personality differences among users. 
Task types were found to affect users’ intentions to use secretive behavior, primarily through influencing perceived external judgment regarding LLM usage. 
Our results yield important insights for future work on designing interventions to encourage more transparent disclosure of the use of LLMs or other AI technologies.
\end{abstract}

\ccsdesc[500]{Security and privacy~Human and societal aspects of security and privacy}
\ccsdesc[500]{Computing methodologies~Discourse, dialogue and pragmatics}
\ccsdesc[500]{Human-centered computing~Human computer interaction (HCI)}

\keywords{Large-language models (LLMs), AI transparency, Privacy, Mixed-Methods Studies}


\maketitle

\section{Introduction}

The advancements in Large Language Models (LLMs), coupled with the wide availability of LLM-based conversational agents, have democratized the power of AI.
Given the open-ended nature of LLMs, the purposes, contexts, and consequences of using them become more directly determined by the end-users.
\revision{As a result, users are playing an increasingly important role in conforming to responsible AI principles such as fairness, transparency and privacy~\cite{diaz2023connecting, dignum2023responsible}, and ensuring AI is not used in a way that causes harms to humanity.}

Transparency, as one of the responsible AI principles, has been significantly impacted by LLMs~\cite{liao2023ai}.
\revision{Prior literature on AI transparency has focused on transparency via explainability techniques, regulatory or legal frameworks, or data-usage disclosures, attributing the responsibility of providing transparency to AI practitioners, researchers, or other technical experts~\cite{corbett2023interrogating}.
However, the democratization of AI, especially LLMs, has extended this responsibility to \textit{AI users}, raising transparency requirements regarding AI usage.}
Due to the potential issues in LLM-generated content, including bias~\cite{abid2021persistent, isaza2023fairy}, hallucination~\cite{bender2021dangers, dahl2024large, yao-etal-2024-samples}, invasion of intellectual property~\cite{karamolegkou2023copyright}, and threats to human autonomy~\cite{parasuraman2010complacency}, it has become crucial for everyday users of LLM-based services to honestly and accurately report when and how LLM-generated content is used in real life.
This is not only important for cautioning against and preventing potential negative consequences caused by LLMs, but also serves as a prerequisite for studying and mitigating these negative impacts.
Disclosure of the use of LLMs has become a basic requirement for \textit{AI users} in various domains to adhere to, such as academic research~\cite{ACM2023AuthorshipPolicy, Elsevier2024AIWritingPolicy}, education~\cite{Harvard2024ChatGPTGuidelines, UMNLib2024LibGuide, UBCLibrary2024CitingGenAI, MSU2024GenAIGuidance}, freelance jobs~\cite{Upwork2024ChatGPTStance}, and online publishing~\cite{Medium2024DistStandards}.
For example, academic publishers like ACM~\cite{ACM2023AuthorshipPolicy} and Elsevier~\cite{Elsevier2024AIWritingPolicy} permit the use of generative AI such as ChatGPT but require full disclosure on how AI has been used for creating content in the publications, given that a lack of transparency in such instances might result in overlooked issues concerning the scientific rigor of the published work~\cite{kaebnick2023editors}.

On the other hand, a recent study~\cite{zhang2024s} on LLM-based conversational agents discovered a novel privacy concern focusing on the users' apprehension that \textit{the action of using LLM-based conversational agents} in a certain context may be discovered.
For example, the paper documented a case in which a non-native English speaker used ChatGPT for revising emails and was worried that her professors might ``change their attitudes to her'' once discovering her reliance on AI and hoped ``they will never know I used AI to do that (write emails).''
This fear could ultimately result in a tendency to hide the use of AI and keep it a secret from others, suggesting a general problem concerning the tension between privacy needs of the individual and transparency needs for the greater good.

However, aside from a few examples, \textbf{there is still a lack of systematic understanding of the landscape, mechanisms, and implications of secret use of LLMs}. Without such understanding, it is infeasible to gauge its potential impact, identify mitigation strategies when privacy and transparency are at odds with each other, and design effective interventions to encourage transparent disclosure of LLM use.
To bridge this gap, we aim to thoroughly study the phenomenon of the secret use of LLMs.
We seek to answer a fundamental question: \textit{When and why do people hide their use of LLMs\footnote{In this paper, we define the scope of ``use of LLMs'' to cover a wide range of activities that involve using content generated by LLMs in real-world tasks. The interaction paradigm might involve widely-used LLM-based conversational agents such as ChatGPT, as well as those used by tech-savvy users, such as open-source LLMs like Llama. We have both types of users in our sample.}?} 

We took a mixed-methods approach to answer this research question, using two sequential studies. 
Given the limited examples found in prior research, we first carried out an online survey with \surveyallusernumber{} LLM users to gain a holistic understanding of the landscape of the phenomenon. 
Through qualitative analysis of \surveycasenumber{} real-world cases of secret LLM usage reported by the users, we summarized the contexts in which secret use of LLMs occurred and developed a typology of causes of users' secret use of LLMs.
We observed that secret LLM usage happened in critical domains such as academic writing and work tasks, and the causes can mainly be divided into two categories -- how users view their usage of LLMs (internal judgement), and how users perceive others view it (perceived external judgement).

Informed by the survey results, we conducted a controlled experiment study with 300 LLM users to quantitatively measure the variance of secret LLM usage among different types of tasks, and investigate the mechanism behind the secret LLM usage behavior.
We focused on understanding the factors influencing the \textit{intention to perform the secret use behavior} at two levels: \revision{passive non-disclosure, where users do not actively disclose their use of LLMs but are honest if asked, and active concealment, where users deliberately hide their use of LLMs even when being directly inquired about it.}
Compared to the baseline task, ``General Information Search,'' we found that the five experimental tasks - ``Creative Writing,'' ``Academic Writing,'' ``School Works,'' ``Work Tasks,'' and ``Social Connection'' - all resulted in significantly higher intentions of both passive non-disclosure (40\% vs. 72\%) and active concealment (16\% vs. 54\%).
However, we did not observe a significant impact due to individual differences, which suggests that the secret use behavior generally applies to people with different demographic backgrounds and personality traits.
Our mediation analysis further reveals that task types primarily affect the intention to use secretive behavior by influencing perceived external judgment.
This provides important insights for future work on designing interventions to encourage more transparent disclosure of LLM/AI use.

With the findings of the two studies, we further discuss the societal and individual-level implications of the secret use of LLM behaviors.
This includes the different priorities between privacy and transparency in various contexts, as well as the potential emotional stress caused to the user by the secretive behavior.
We conclude with a set of suggestions for stakeholders to build community norms and improve community cultures, in order to mitigate the barriers to transparent disclosure caused by perceived external judgement.
We also identify future research directions to further explore this topic.

\section{Background and Related Work}

\subsection{Secretive Behavior}

We consider the secret use of LLMs one special case of secretive behavior, and review the psychology literature on secretive behavior in other contexts.
A broader view of secrecy regards it as an intention to keep information unknown to one or more one individual, with two main psychological paths~\cite{slepian2022process, slepian2017experience}. 
One is ``mind-wandering to secret,'' an internal and subconscious process where secret occupies an individual's mind~\cite{slepian2022process, slepian2017experience}. 
The other is ``concealment of secret,'' an external process where one makes conscious decisions to not disclose the information to others~\cite{slepian2022process, slepian2017experience}.
Our study falls under the path of ``concealment of secrets'', focusing on the secretive behavior or intentions that associate with their external actions.

Several studies on secretive behavior lies in secret consumption in different areas, such as secrete consumption on counterfeit~\cite{bian2016new}, secret consumption of meat~\cite{khara2020we} and secret snack consumption behavior~\cite{forbes2016analysis}.
A large portion of studies about specific secretive behavior lies in area of ``knowledge hiding behavior''~\cite{connelly2012knowledge, serenko2016understanding, arshad2018workplace, kumar2018you}.
\citet{connelly2012knowledge} and \citet{serenko2016understanding} introduced early research that formally studied the factors influencing people's knowledge hiding behavior in organizations. 
They used a mixed-methods approach, asking participants to share their personal related stories and then measuring participant's perceptions related to the behavior~\cite{connelly2012knowledge, serenko2016understanding}.
These studies inspired our study design, which also employed a mixed-methods approach, including an open-ended case-listing survey followed by a controlled online experiment.

Research on users' secret use of AI or LLMs is limited, showing a gap that our study aims to address. 
One related study about AI Ghostwriter Effect~\cite{draxler2023ai} found that users refrain from declaring AI authorship even though they do not consider themselves the owners and authors of AI-generated text.
Their study design focused on an AI-assisted postcard writing task, while our research examined tasks that could broadly engender passive non-disclosure and active concealment behaviors.
Through the unique lens of secret use behaviors, our work identifies new social factors that affect these behaviors, such as users' own moral judgments and their perceived social expectations.

\subsection{AI Transparency}

With the deeper integration of AI into people's daily lives and decision-making, the lack of transparency in AI systems can lead to doubts about the reliability of decisions made by AI~\cite{von2021transparency, larsson2020transparency, toy2023transparency, chander2018working}. 
Current research in the field of HCI aimed at enhancing AI transparency can be divided into two primary categories: the ways in which users perceive these AI systems, and the development of tools by researchers and engineers to increase the explainability of AI systems.

Researchers conducted lots of studies to understand how users perceive AI systems \cite{rzepka2018user, shin2020user, kocielnik2019will, shin2021effects}.
\citet{shin2021effects} found that users tend to view algorithms as more trustworthy and useful when they perceive them to be fair, accountable, transparent, and explainable.
Similarly, \citet{yu2022artificial}, through interviews with 235 employees about their views on AI at work, found that transparency in AI decision-making improved both its perceived effectiveness and addressed the associated discomfort.

Given the insights about users' perceptions and needs, researchers have developed tools to enhance users' ability to assess the transparency of the AI systems they use \cite{zhang2020lay, lundberg2019explainable, hu2021xaitk, karim2023explainable, dwivedi2023explainable}.
For example, Transparency-Check~\cite{schelenz2023transparency} is a tool to aid users in evaluating the transparency of AI-based systems, particularly those with personalization.
This tool includes questions around data collection, algorithmic models, personalized recommendations, and user control.

Prior research of AI transparency has primarily viewed AI researchers and practitioners as the information provider, with users being the recipients of the information~\cite{corbett2023interrogating}.
However, the user-driven, open-ended nature of the use scenarios of large language models (LLMs) and the extensive use of LLM-based conversational agents in critical domains have raised concerns about the transparency of how LLMs are used by users, anticipating users to play an important role as information providers as well.
By examining the contextual factors and psychological process that affect users' secret use behavior, our findings shed light on users' ability to fulfill their responsibility towards AI transparency.
Our research broadens the scope of research in the field of AI transparency.

\subsection{Existing Community-Level Requirements of Transparency for LLM Use}
\label{sec:community-level-requirements-of-transparency}
With the rapid growth of LLMs, some agencies have announced regulations or recommendations regarding the use of LLMs in their respective communities. 
Despite their different strategies of handling the impact of LLMs, a common theme among these regulations and recommendations is transparency.
We summarize major examples below.

\paragraph{Universities} 
Universities have implemented varied regulations regarding students' obligations to disclose their use of LLMs in coursework.
For example, University of Minnesota requires students to reference all the content in their assignment generated by the AI tools, or they will be regarded as scholastic dishonesty~\cite{UMNLib2024LibGuide}.
University of British Columbia allows students to use LLMs with instructors' approvals and provides a guide on how to cite contents generated by LLMs~\cite{UBCLibrary2024CitingGenAI}.

\paragraph{Journalism, Social Media and Blog} 
Journalism and social media platforms also put regulations to standardize the use of LLMs.
The Associated Press encourages its journalists to verify that material received from other sources is also free from AI-generated content, which relies on a reliable source of whether AI is involved in the generation of the content~\cite{AP2024ChatGPTStance}.
Medium, an online publishing platform, announced that the use of LLMs is permitted for writing stories~\cite{Medium2024DistStandards}.
However, to promote transparency, stories that incorporate AI assistance must be clearly labeled as such.
YouTube encourages its content creators to maintain transparency regarding the use of AI~\cite{YouTube2024ResponsibleAI}.

\paragraph{Academia}
Academia conferences and journals often request authors to fully reveal their use of LLMs. 
For example, the Association for Computing Machinery (ACM) and Elsevier have updated their policy regarding authorship and the use of LLMs~\cite{ACM2023AuthorshipPolicy, Elsevier2024AIWritingPolicy}.
The creation of content with LLM tools and technologies is permitted, but it must be fully disclosed in the work.

\paragraph{Intergovernmental and governmental organization}
Regulations requiring the disclosure of LLM use are evolving globally and vary across regions and specific applications. As an example,
The Canadian federal government has released new guidelines for employees who wish to use artificial intelligence tools like ChatGPT in their work~\cite{Canada2024GenerativeAI}.
The guidelines require employees to clearly identify content generated by LLMs and to inform users when they are interacting with an AI tool.

\paragraph{Workplaces}
Companies and platforms are increasingly concerned about their employees' use of LLMs. 
For instance, Upwork, a freelancing platform, mandates its freelancers to explicitly disclose to clients whenever artificial intelligence is used in content creation~\cite{Upwork2024ChatGPTStance}. This includes job proposals and direct messages.
Additionally, an article titled ``Writing A ChatGPT Policy For Your Organization'' published on LinkedIn advises employers to establish a LLMs policy for their companies, emphasizing the importance of transparency in AI use in the workplace~\cite{LinkedIn2024ChatGPTPolicy}.

\section{Study 1: Online Survey}

To understand when and why people hide their use of LLMs, we first conducted an online survey with \surveyallusernumber{} LLM users, asking them to reflect on their past experiences of secret use of LLMs, list the secret use scenarios, and explain the reasons.
\surveyconcernedusernumber{} participants (62\%) reported they had LLM secret use experience while others not. 
Finally, we collected \surveycasenumber{} unique and valid secret use cases with rationales (noting that some submitted multiple cases).
We qualitatively analyzed the data and derived a taxonomy summarizing eight common reasons that lead to the secret use.

\subsection{Survey Design}
The main task in the questionnaire was to share the secret use experience of LLMs.
At the beginning of the survey, there were screening questions to determine whether participants had relevant experiences with LLMs and any secret usage.
Given that users may not be familiar with jargon like ``Large Language Models (LLMs)'' or ``LLM-based conversational agents''~\cite{zhang2024s}, we first explained the qualitifications in the screening process by showing example LLM applications such as ChatGPT, OpenAI API Playground, Bard, Pi.ai, and Claude.ai.
We then refer to them throughout the survey using a term more friendly to lay users, ``AI chatbot''.
Responses from those without LLM experiences were excluded in our results.

\begin{figure}
    \centering
    \includegraphics[width=0.7\linewidth]{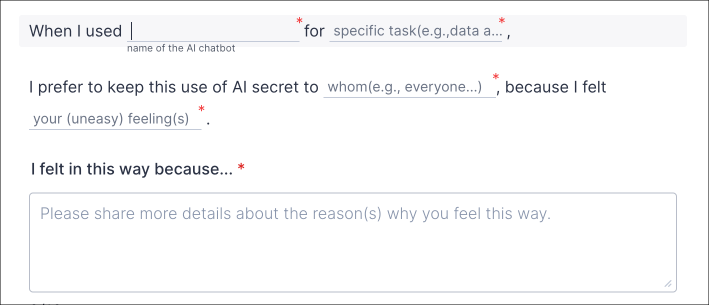}
    \caption{Fill-in-the-blank questions for participants to share their experiences of secret use of LLMs}
    \label{fig:survey-blank-filling-example}
\end{figure}

For respondents who reported no secret use of experience or intention on the secret use, we asked the follow-up questions about their reasons.
Respondents with secret usage experience were prompted to share up to three stories one by one detailing their scenarios, reasons for secrecy, and coping strategies.
To ensure that the users provide precise and comprehensive information to describe the scenario, we used fill-in-the-blank questions.
These questions guided them to reflect and report on each case in various aspects, including the name of the LLM service, the specific task, the entities they want to keep the AI use secret from, and their feelings.
This was followed by three questions on reasons for secrecy, feelings about secret usage despite negative feelings, and any measures taken to maintain secrecy (\autoref{fig:survey-blank-filling-example}).

The survey questionnaire was built on Jotform\footnote{\href{https://www.jotform.com/}{Jotform} is a website for building online survey forms.} and distributed through Prolific\footnote{\href{https://www.prolific.co}{Prolific} is a website for recruiting research study participants.}.
Our pilot study indicated it took about 4 minutes to complete the survey if the participant sharing only one secret use case.
Therefore, basic compensation was set at \$0.4, with an additional \$0.2 for each extra case shared.
To ensure a diverse sample, we released the survey at different times across weekdays and weekends, and controlled for a balanced gender distribution between female and male participants.

\subsection{Qualitative Analysis Methods}
We conducted open coding to analyze the scenarios of the secret use of AI reported by the participants, as well as the self-reported feelings and rationales.

\subsubsection{Reason of Secret Use Analysis}
Our main goal was to establish a typology of reasons why participants wanted to hide their usage of LLMs.
To this end, two researchers iteratively coded the self-reported feelings and rationales for each case reported by the participants.
The two researchers first collectively reviewed a subset of the valid cases to establish an initial coding scheme.
Next, they carried out several rounds of independent relabeling, each time with a small subset of the dataset, in order to refine the typology and align the two coders' interpretations of the categories.
At the end of each round, they calculated the inter-rater reliability (IRR) and discussed resolving discrepancies and adjusting the coding scheme when needed.
This process was repeated iteratively until a high IRR (\(> 0.8\)) was reached. 
The coding scheme was finalized after five rounds of coding and revisions.
To validate the coding scheme, we collected a new dataset using the same protocol.
The same two coders independently coded the new dataset.
The inter-rater reliability (IRR) score on the validation set was 0.834, as measured by Krippendorff's alpha. 
This measurement is a standard for inter-rater reliability measure for codes that are not mutually exclusive~\cite{hayes2007answering}.
We selected this measure because each case may be associated with multiple reasons.

Our final coding scheme comprised nine categories, eight for specific reasons and one category ``Other or unclear reason'' for labeling short or ambiguous responses that did not provide enough information for analysis.

\subsubsection{Scenario of Secret Use Analysis}

We conducted a thematic analysis to analyze the scenarios reported by the participants.
Two researchers independently reviewed all the valid cases, and each identified some themes around the scenarios.
They then discussed together to combine the two sets of themes and synthesize them into a list of themes characterizing the tasks that engendered the secret use behavior.
Six themes about the scenarios were identified in the final coding scheme.
We did not use IRR because the primary goal is to yield themes of scenarios rather than seeking agreement~\cite{mcdonald2019reliability}.

\subsection{Methodological Limitations}
Our method has certain limitations that should be taken into account when evaluating the results.
Firstly, the information in the participants' responses is often limited, which may not fully capture their reasoning.
To facilitate data collection, we asked participants to describe their feelings by filling in the blanks in a questionnaire, as shown in \autoref{fig:survey-blank-filling-example}.
Participants tended to use only one or two sentences to explain their feelings.
Consequently, when coding the data, we noticed that a single response could be interpreted in multiple ways.
To address this, we chose to interpret the response in a conservative way, and conducted several rounds of iteration to carefully define our coding scheme and strive for as much consensus as possible. 
Secondly, the scenarios identified from our sample are not exhaustive.
The cases reported by users should be considered common cases, but they may not cover all the secret use of LLMs in real life.
More uncommon yet critical cases might remain to be explored by future research using other research methods.

\subsection{Survey Results}

\subsubsection{An Overview of the Contexts of Secret Use of LLMs}
\label{sec: survey-context-results}
Based on participants' reports, we identified the following types of scenarios that have led to a tendency to conceal the use of LLMs, including \textit{creative writing} (11\%), \textit{academic writing} (3\%), \textit{school works} (10\%), \textit{work tasks} (15\%), \textit{social connections} (14\%), \textit{sensitive topics} (36\%), and other miscellaneous cases\footnote{We did not notice these cases showing a recurring pattern, so decided to group them together.} (including entertainment, brainstorming, etc).
We introduce each scenario in depth below.

\paragraph{Creative writing}
This category referred to scenarios where participants used LLMs for creative writing and prefer to keep the LLM usage secret from other people.
For example, SP15 (P15 in survey study) and SP25 used ChatGPT for writing fiction and fanfiction, and chose to keep their use of AI tools a secret.
Similar secretive behavior was obeserved in SP36, who used NovelAI for writing stories, and SP56, who used ChatGPT for YouTube video scripts creation.
This category was specifically created to explore how people feel when they combine AI-powered tools with their own creativity.

\paragraph{Academic writing}
Several participants expressed a desire to conceal their use of LLMs in academic writing, such as writing papers or articles and generating research ideas. 
For example, SP8 used ChatGPT for paragraph writing in academic works and chose to keep the usage secret from everyone, believing that the ideas generated were not originally his.
We designated this category to gain insights into how people employ LLMs in conjunction with their expertise in academic fields.

\paragraph{School works}
Participants also indicated a desire to conceal their use of LLMs in school-related tasks, such as completing an essay or homework assignments.
Comparing with ``academic writing'', the ``school works'' category places a greater emphasis on the educational environment.

\paragraph{Work tasks}
Many participants mentioned using LLMs for various work-related tasks, preferring to keep this usage confidential.
Examples including coding, data analysis, and other general job responsibilities.
The types of tasks are closely related to participants' work domains and job positions.
For example, SP16 used ChatGPT for policy writing but was concerned about being discovered by his employer.
SP13 used ChatGPT to develop marketing plans and preferred to keep this use a secret within his industry.

\paragraph{Social connection}
A large portions of reported secret use experiences involved using LLMs to maintain social connection including personal and professional relationships.
For example, SP17 and SP40 used ChatGPT to generate love messages to their partners and kept the use of LLM tool secret.
SP94 chose to hide their use of Bard in writing work emails from colleagues.

\paragraph{Sensitive topics}
Tasks under this category are mostly related to sensitive interaction content, such as sex or erotica, health, medical topics, personal advice or therapy.
For example, SP54 used Character.ai for sex roleplay and was afraid of being found out.
SP84 turned to Bing chat for medical advice and thought that his medical details were not free to other. 
This category often associates with the reason \revision{``Content sensitivity concerns''} in \autoref{sec: survey-cause-results}, where people are more worried about the sensitive content they interact with using LLMs, rather than the mere usage of AI itself.

\subsubsection{A Typology of Causes to Secret Use of LLMs}
\label{sec: survey-cause-results}
We then introduce the results about the reasons behind the secret use of LLMs, synthesized from the self-reported reasons.

We first want to introduce a special category, \revision{``Content sensitivity concerns''}, which emphasizes concerns related to the task itself or the content shared via the interaction with LLMs being discovered, rather than the fact of using LLMs. 
For instance, SP107 consulted ChatGPT for health advice and wanted to keep the interaction private: ~\blockquote{My health and health record are private things. When I go online in search of information on my condition or a potential condition it's important to keep that confidential.}
Since this category diverges from our research focus on the concealment of LLM usage, we have excluded it from our detailed analysis.

After excluding the categories of ``Other and unclear reasons'' (11 cases) and ``Content sensitivity concerns'' (37 cases), we identified seven categories encompassing 77 valid instances directly related to concealing LLM usage. 
These categories are grouped into two themes: \textbf{Internal Judgment} and \textbf{Perceived External Judgment}.

Internal judgment encompasses reasons related to the participants' self-assessment of using AI. This includes \textit{``Questioning their own competence''}, \textit{``Not a wise choice to use LLMs''}, and \textit{``Moral doubts on using LLMs''}.

\paragraph{Questioning their own competence (8/77, 10\%)} 
When using LLMs for completing tasks, some individuals experienced feelings of incompetence, inadequacy, or unprofessionalism.
For instance, SP3 felt incompetent when using ChatGPT for coding assistance, stating,~\blockquote{It's not the best feeling to know you need help from an AI bot.}
Similarly, SP7, who used ChatGPT for creating an essay outline, felt ashamed and preferred to keep her use of LLMs a secret.
She expressed,~\blockquote{I think it would make me feel inadequate because I used an outline.}

\paragraph{Not a wise choice to use LLMs (9/77, 12\%)} 
Some individuals frequently questioned whether using LLMs was the effective or optimal choice for completing tasks.
In these cases, participants often implied a doubt on the effectiveness of the LLM-based CAs.
For example, SP37 used ChatGPT for family advice while she felt ashamed and preferred keep this usage secret as she said~\blockquote{I know that the AI doesn't understand family matters and how to respond in a way that a human would.}
SP31 kept her usage of ChatgGPT in health consulting a secret and she said~\blockquote{It's stuff I should of went to a doctor for and not been asking a chatbot.}

\paragraph{Moral doubts on using LLMs (26/77, 34\%)} 
While using LLMs, some individuals believed that employing these tools for certain tasks might be immoral.
For example, SP66, who utilized ChatGPT to help draft a recommendation letter for a National Interest Waiver, felt like a fraud towards the people she worked with.
Similarly, SP7 felt he was cheating by writing with the assistance of ChatGPT, stating,~\blockquote{It felt like I was cheating. I used it (ChatGPT) to write when I should be able to do that on my own.}

A different category of reasons, perceived external judgment, covers reasons where one conceals their use of LLMs due to apprehension about external criticism.
This might have stem from the general societal ambivalence towards emerging, potent technologies like LLMs.
This category included four reasons, \textit{``Fear of capabilities being critiqued''}, \textit{``Fear of personality/attitudes being judged''}, \textit{``Sincerity concerns in personal relationships''} and \textit{``Using AI for certain tasks is considered unethical/inappropriate by others''}.

\paragraph{Fear of capabilities being critiqued (8/77, 10\%)} 
Some individuals chose to conceal their use of AI due to fears that their abilities would be judged by others. 
For instance, SP85, who used ChatGPT for resume writing, mentioned that,~\blockquote{If I am using an AI chatbot for help with writing my resume, some people might think I am inadequate.}
Similarly, SP88 employed ChatGPT for email writing and kept it hidden from her boss, feeling uncomfortable with her method.
She stated,~\blockquote{I didn't want them to think that I was too dumb to write an email myself.}

\paragraph{Fear of personality/attitudes being judged (15/77, 20\%)} 
Some users are concerned about potential judgment of their personality and attitude.
For example, SP3 used ChatGPT to ask for personal advice and chose to keep it secret.
She described that~\blockquote{People could look at me oddly for using the bot or maybe taking it's advice.}

We specifically identified \textit{Sincerity concerns in personal relationships (7/77, 9\%)} as a special case of this category, where individuals were concerned that using AI for personal interactions, such as writing a love message, might be perceived as insincere and could negatively impact their relationships.
These cases are noteworthy because they highlight the delicate balance between technological convenience and the authenticity of human emotions in personal communications.
P75, for example, used ChatGPT to write emails but preferred to keep this fact hidden from the receiver.
He explained,~\blockquote{I felt that they would think I don't care about the answer. People are sensitive when they realize an AI replied to their emails and think the person that sent the email doesn't care to write their own email.}

\paragraph{Fear of AI use being considered unethical/inappropriate by others (6/77, 8\%)} 
Another reason deterring people from disclosing their use of LLMs is the belief that ``Using AI for certain tasks is considered unethical/inappropriate by others.''
This is distinct from the category where individuals have ``Moral doubts on using LLMs,'' which relates to users' own internal judgment.
The concern about ethical judgments is rooted in the anticipation of others' moral standards, rather than one's own.
For instance, SP96 used LLaMA for creative content generation and kept the usage secret to his potential audiences because he felt the usage would be disapproved.
He stated,~\blockquote{People in the audience I was targeting at the time are predisposed to mistrust technology, particularly AI.}

\section{Study 2 Experiment: Hypotheses}

Our first survey yielded qualitative insights around the scenarios in which users hide their use of LLMs and the underlying reasons for such behavior.
Based on the survey results, we conduct a controlled experiment to quantitatively investigate the effect of task types and individual differences on the propensity for covert LLM usage, and the mediation effects of users' internal and perceived external judgments.
We formally introduce the hypotheses below.

\subsection{Intentions of Secret Use of LLMs}
Previous research has identified two primary forms of secrecy behaviors: concealment by an act of omission (withholding information) and concealment by an act of commission (actively deceiving or providing false information)~\cite{pennebaker1989confession, lane1995cognitive, kelly2002psychology, bok2011secrets}. 
In our context, the secret use of LLMs may also be exhibited in these two forms.
Therefore, to provide a more comprehensive understanding about factors that affect the intentions of secret use of LLMs, we measure two variables representing different levels of concealment behavior in our experiment: passive non-disclosure and active concealment.
In our study, we defined \textbf{passive non-disclosure} as not actively disclosing the use of LLMs while being honest about it when asked.
And \textbf{active concealment} refers to the behavior of actively hiding the use of LLMs, even when directly inquired about it.

\subsection{Effects of Task Types on the Intentions of Secret Use of LLMs}
Our qualitative results about the scenarios of secret use of LLMs (see \ref{sec: survey-context-results}) suggest that certain types of tasks are associated with individual users' secret use of LLMs.
This observation leads to the hypothesis that the intention to conceal the use of LLMs may be influenced by the nature of the tasks users are engaged in.
Furthermore, related studies and theories emphasize the basic role of context in shaping people's intention to perform certain behaviors~\cite{bommer1987behavioral, leonard2004influences, kreie2002copyright, hoffman1998women}.
For example, the ethical decision-making model introduced by~\citet{bommer1987behavioral} highlights context as a fundamental element influencing intentions to perform ethical behaviors, and even modulate the impact of other variables in diverse ways~\cite{leonard2004influences, kreie2002copyright, hoffman1998women}.
Similarly, Goffman's theory of self-presentation considers the task people perform as foundational elements when discussing people's ``frontstage'' and ``backstage'' behavior~\cite{goffman2016presentation}.
The psychology of secrecy and contextual privacy further states that people decide how to perform self-disclosure behaviors depending on the situation~\cite{slepian2022process, nissenbaum2004privacy}.
Drawing upon these related theoretical frameworks and our survey findings, we hypothesize that:

\textbf{H1} Different task types cause different levels of intentions to hide the use of LLMs.

\subsection{Effects of Individual Differences on the Intentions of Secret Use of LLMs}
Prior studies have indicated the potential influence of individual differences on the intention of secrecy~\cite{slepian2022process, leonard2004influences}.
In our experiment, we build upon the findings from the first survey and prior literature to establish hypotheses about how individual differences influence individual users' intentions of secret use of LLMs.

\paragraph{Self-esteem}
Survey results about causes of secret use of LLMs introduced two main categories of self-reported reasons: internal judgment and perceived external judgment (see \ref{sec: survey-cause-results}). 
The way individuals construct these judgments and their response to them, which in turn influences their behavior, can be potentially affected by their level of self-esteem~\cite{leary2000nature}. 
For example, individuals with high self-esteem may feel more confident and less concerned about external judgments, potentially making them more open to disclosing their use of LLMs. 
Therefore, we hypothesize that:

\textbf{H2.1} Users' level of self-esteem affects their intentions to hide the use of LLMs.

\paragraph{Self-efficacy}
Judgments about users' capabilities are featured in the self-reported reasons for concealing the use of LLMs (see \ref{sec: survey-cause-results}), encompassing both internal (questioning users' own competence because of the need of AI) and external perspectives (fear of being critiqued by others for using AI assistance).
These types of judgments are intrinsically linked to an individual's self-efficacy~\cite{bandura1977analysis}. 
For instance, individuals with higher self-efficacy might feel confident in their abilities, reducing the need for concealment of AI usage. 
Therefore, we hypothesize that:

\textbf{H2.2} Users' level of self-efficacy affects their intentions to hide the use of LLMs.

\paragraph{Prosocialness}
Prosocial users, who are more inclined towards societal welfare~\cite{caprara2005new}, may view the disclosure of their LLM usage as a contribution to the ``greater good'' in promoting responsible AI usage. 
However, they might also balance this against potential ethical judgments from others. 
This interplay suggests that one's level of prosocialness could be a potential factor influencing their decision-making process regarding the disclosure of AI usage.
To explore this, we hypothesize that:

\textbf{H2.3} Users' level of prosocialness affects their intentions to hide the use of LLMs.

\paragraph{General privacy concern}
General privacy concerns often revolve around the control and unauthorized use of personal data~\cite{malhotra2004internet}.
While the concerns about being discovered using LLMs do not directly pertain to data privacy, it relates to the revelation of the fact that AI is being used. 
Recognizing this nuanced aspect of privacy, we are interested in exploring how general privacy concerns impact the intention to hide the use of LLMs. 
Hence, we hypothesize that:

\textbf{H2.4} Users' level of general privacy concern affects their intentions to hide the use of LLMs.

\paragraph{Demographics}
A large body of research has identified the influence of demographic factors on people's intentions to perform secrecy behaviors~\cite{slepian2022process}, ethical technology use behaviors~\cite{leonard2004influences}, and technology adoption in the context of tensions between ``the greater good'' and ``individual privacy''~\cite{li2021makes, nissenbaum2020privacy}.
For example, gender and age have been shown as predictors of ethical or unethical behavioral intention~\cite{dawson1997ethical}. 
Exploring the effects of demographics on the intentions of LLM secret use can help propose targeted interventions for specific population groups. 
Therefore, we form the following hypothesis to investigate this possibility:

\textbf{H2.5} Users' demographic factors (gender, age, LLM use frequency, and education) affect their intentions to hide the use of LLMs.

\subsection{Mediation Effects of Users' Internal Judgement and Perceived External Judgement about the LLM Usage}
The reasons users self-reported for concealing their use of LLMs fundamentally reflect their perceptions regarding the secret usage behavior (internal judgement and perceived external judgement), as discussed in \autoref{sec: survey-cause-results}.
According to the Self-Discrepancy Theory, when there are discrepancies between a person’s actual self (e.g., using LLMs for a task) and their ideal or ought self (e.g., their negative internal judgment or perceived external judgement), negative emotions are elicited.
This can significantly influence people’s behaviors~\cite{orellana2000decisional, mandel2017compensatory}.
Following the investigation of the impact of each independent variable, we want to further understand the causal mechanisms behind the relationship between the independent variables and the outcome variable - the secrecy behavior.
Specifically, we focus on the mediation effect of the internal judgement and perceived external judgement, to examine how task types and individual differences influence individuals' intentions to hide the use of LLMs through these two categories of internal beliefs. 
More formally, we propose the following hypothesis:

\textbf{H3.1} Internal judgement mediates the relationship between the independent variables (i.e., task types and individual differences) and the intentions to hide the use of LLMs.

\textbf{H3.2} Perceived external judgement mediates the relationship between the independent variables (i.e., task types and individual differences) and the intentions to hide the use of LLMs.

\section{Study 2 Experiment: Methodology}
\label{sec:experiment}

To test the hypotheses about factors that influence individual users' intention of secret use of LLMs, we conducted a randomized between-subjects online experiment ($N=300$), using Prolific for recruitment.
\revision{The sample size was determined before the formal study based on the power analysis results ($\alpha=0.05$, power > 0.8, the effect size $f^2=0.03$).
The effect size \(f^2\) was estimated using pilot study data ($N=76$), indicating a small to medium impact according to Cohen's criteria~\cite{cohen2013statistical}.}
The questionnaire was hosted on Qualtrics\footnote{\href{https://www.qualtrics.com/}{Qualtrics} is a website for building online survey study.}.
The experiment was conducted in December 2023 and January 2024.
Our study has been reviewed and approved by our institution’s IRB.

\subsection{Experiment Design}
Our experiment followed a between-subjects design, in which participants were randomly assigned into six conditions, including a baseline condition ``General information search'', as well as five experimental conditions based on the five task themes identified from our first study (\autoref{sec: survey-context-results}).
The selection criteria for the baseline condition include that 1) it is a common LLM use case~\cite{zhang2024s}; 2) it was not mentioned as a secret use case by Study 1 participants.
Each condition presents different hypothetical scenarios where users have used LLMs for different tasks.
The task description shown to the participants can be found in \autoref{sec: appendix-scenario-description}.
These manipulations allow us to test the effects of task types on individual users' intentions to conceal their use of LLMs (H1). 
Then we gathered data on participants' self-reported perceptions (internal and external judgements) of LLMs based on the hypothetical scenario.
This approach allows us to explore the mediation effects of users' certain perceptions about LLM usage (H3.1 - H3.2).
Participants were finally asked to fill out validated scales and provide demographic information to facilitate testing of H2.

\subsection{Experiment Procedure}
\label{sec:experiment-procedure}
Our experiment consisted of three steps.

\textbf{Step 1.} Task description and intention questions: 
Participants first received a hypothetical scenario describing the use of LLMs for tasks, randomly selected from the six conditions (one baseline, five experimental).
They were then asked to report their intentions to conceal this usage within the task context.
We clarified that the concealment intentions referred specifically to the fact of using LLMs, not the data transmitted during the interactions.
After reporting the intentions, they were asked to elaborate on their rationales, which helps us understand their selections.

\textbf{Step 2.} Questions about the LLM usage: 
Participants were re-presented with the same hypothetical scenario and responded to two sets of questions about their internal and external judgements of LLM usage in that context.
To minimize order bias, question order was randomized at both the group level and within each group.
We inserted an attention check question after all other questions (``Please select [Strongly agree] to show you are paying attention to this survey.''). 
If they selected other answers, their responses would be excluded from our final dataset.
In the second page, participants were asked whether they have similar experiences. 
For those who had similar experiences, an open-ended question would be provided at the next page that allowed participants to freely share their similar experience, choice on the concealment, and the reasons. 
To prevent users' responses from being influenced by later questions, turning back to the previous pages was not allowed at any stage of the whole questionnaire.

\textbf{Step 3.} Questions about individual differences: 
In this final step, participants completed validated scales measuring self-esteem~\cite{rosenberg1965rosenberg}, self-efficacy~\cite{schwarzer1995generalized}, prosocialness~\cite{caprara2005new}, and general privacy concerns~\cite{malhotra2004internet}, presented in a randomized order across different pages. 
This was followed by demographic questions (age, gender, and education background).

\subsection{Operationalization}

\subsubsection{Dependent Variables}
\label{sec:experiemnt-operationalization-dep-var}
We asked participants to report their intentions on two forms of concealment on the use of LLMs on a 7-point likert scale (1 = strongly disagree, 7 = strongly agree) in Step 1 (\autoref{sec:experiment-procedure}).
After introducing the hypothetical scenario, we asked participants to rate to what extent they agreed or disagreed with the following statements:
\textbf{Passive non-disclosure: }``I would not mention that I used an AI chatbot for this task/purpose on my own''.
\textbf{Active concealment: }``Even when others ask me, I prefer to hide the use of an AI chatbot for this task/purpose''.

\subsubsection{Independent Variables}
We introduce the operationalization of the two categories of independent variables, task types, and individual differences.

\textbf{Task type:} 
The task type is a categorical variable with six levels, as six task types were selected to create hypothetical use scenarios of LLMs.
We chose the ``General information search'' task type, which is a common use case of LLMs~\cite{zhang2024s} and not mentioned by our survey participants, as the reference level. 
The other five levels of the variable are ``creative writing'', ``academic writing'', ``school work'', ``work tasks'' and ``social connection'', corresponding to the types of tasks identified from our qualitative analysis of the survey results (\autoref{sec: survey-context-results}).

\textbf{Self-esteem:}
Participants’ self-esteem was measured using the 10-item Rosenberg Self-Esteem Scale with ten questions that are on a 4-point likert scale~\cite{rosenberg1965rosenberg}. 
Higher scores in questions 1, 3, 4, 7, 10 contribute higher self-esteem while other questions contribute inversely~\cite{rosenberg1965rosenberg}. 
To score individuals' self-esteem levels, we first reversed the scores of questions 2, 5, 6, 8, and 9 and then averaged the scores of all questions.
The internal consistency (Cronbach’s alpha) of all 10 questions was 0.93 in our sample, which showed high reliability.

\textbf{Self-efficacy:}
Self-efficacy was measured using a 10-item scale~\cite{schwarzer1995generalized}.
The ten questions are on a 4-point likert scale, and a higher score means higher self-efficacy. 
We define the self-efficacy value for each individual as the average rating of the ten questions.
The internal consistency (Cronbach’s alpha) of all ten questions was 0.91 in our sample, which showed high reliability.

\textbf{Prosocialness:}
We used the 16-item scale to measure participants’ prosocialness~\cite{caprara2005new}. 
The sixteen questions are on a 5-point likert scale, and a higher score means higher prosocialness. 
We define the prosocialness value as the average rating of the 16 questions~\cite{caprara2005new}. 
The internal consistency (Cronbach’s alpha) of all 16 questions was 0.93 in our sample, which showed high reliability.

\textbf{General privacy concerns:}
We used the 10-item Internet Users’ Information Privacy Concerns (IUIPC) scale to measure participants’ general privacy concerns~\cite{malhotra2004internet}. 
The ten questions are on a 7-point likert scale, and higher score means higher privacy concerns. 
We define the general privacy concern value for each individual as the average rating of the 10 questions. 
The internal consistency (Cronbach’s alpha) of all 10 questions was 0.85 in our sample, which showed high reliability.

\textbf{Gender:}
We provided four options for participants to select: ``Male'', ``Female'', ``Non-binary / third gender'', and ``Prefer not to say''.

\textbf{Age:} 
We provided a text input box to allow participants to enter their age.

\textbf{Education:}
We provided 7 options for participants to select: ``Some school, no degree'', ``High school graduate'', ``Some college, no degree'', ``Bachelor's degree'', ``Master's degree'', ``Professional degree'', and ``Doctorate degree''.
We converted the 7 options to integers 1 to 7, with 1 corresponding to ``Some school, no degree'' and 7 to ``Doctorate degree''.

\textbf{Use frequency of LLMs:} \label{sec: using-frequency-question}
We provided eight options for participants to select: ``I've never used any AI chatbots,'' ``I've used it before, but not regularly,'' ``Less than once a month,'' ``Monthly (1-3 times a month),'' ``Weekly (once a week),'' ``Multiple times a week (but not daily),'' ``Daily (almost once every day)'' and ``Multiple times daily (several times throughout each day).''
The option ``I've never used any AI chatbots'' was to exclude participants without experience in using LLMs.
The remaining seven frequency categories were converted into ordinal values, from 1 (I've used it before, but not regularly'') to 7 (``Multiple times daily''), reflecting an ascending order of usage frequency.

\subsubsection{Mediator Variables}
The mediator variables in our experiment are users' internal judgements and perceived external judgements on the use of LLMs.
\revision{We operationalized them as two latent variables, each explained by three statements about the internal judgements or three statements about the perceived external judgements, respectively.
The statements are derived from the reasons for concealing the use of LLMs identified in our first study (\autoref{sec: survey-cause-results}). 
Using latent variables allows for evaluation of item contributions, accounts for measurement error, and provides a more accurate depiction of the relationships between constructs, particularly when the relationships between observed variables and underlying constructs are unknown or complex~\cite{sawatzky2010self, zhang2016privacy}.}
We asked participants to rate their agreement or disagreement with the six statements using a 7-point Likert scale (1=strongly disagree, 7=strongly agree) in Step 2 (\autoref{sec:experiment-procedure}).

\revision{Factor Analysis provides support for our construct validity, showing that two groups of variables have high loadings on their respective factors~\cite{brown2012confirmatory}.
The Cronbach's alpha also indicates high reliability of statements within each group in our sample~\cite{zumbo2007ordinal}.}
The two groups of statements are listed as follows.

\textbf{Internal judgements on the use of LLMs} (Cronbach's alpha = 0.79):
``I feel less competent when I need an AI chatbot to assist me with this task'' \revision{(Factor loading = 0.77)}; ``I have doubts about the effectiveness of using AI chatbots for this task/purpose'' \revision{(Factor loading = 0.68)}; ``I think using the AI chatbot for this task/purpose is immoral or unethical'' \revision{(Factor loading = 0.81)}.

\textbf{Perceived external judgements on the use of LLMs} (Cronbach's alpha = 0.90): 
``Other people will critique my capacities if they know I used the AI chatbot for this task/purpose'' \revision{(Factor loading = 0.88)}; ``Other people will judge my personality or attitudes if they know I used the AI chatbot for this task/purpose'' \revision{(Factor loading = 0.90)}; ``Other people will see using the AI chatbot for this task/purpose as inappropriate or unethical'' \revision{(Factor loading = 0.82)}.

\subsection{Participants}
We recruited participants based in the U.S. through Prolific and compensated \$1.4 each. 
The recruitment process involved several steps to obtain a sample of $N=300$ LLM users, evenly distributed across six conditions (50 participants per group).
First, we included a question about LLM usage frequency (\autoref{sec: using-frequency-question}) to ensure that only LLM users proceeded to the main study.
And then we manually reviewed the responses from the main study to filter out those failing the attention check or providing contradictory responses (e.g., disagreeing with intentions to conceal usage but explaining they would have concealed it in the hypothetical context).
We also verified that the quality of all the open-ended responses in the final sample is high.
As exclusions occurred, we adjusted the randomizer settings to maintain an even distribution across the six conditions.
The recruitment process continued to roll until \experimentvalidsamplesize{} valid responses were reached (50 in each condition).
Meanwhile, we employed the balanced sample distribution mode in Prolific for a balanced gender distribution.
Ultimately, we received \experimentrawdatapoint{} responses, of which 18 were excluded, resulting in a final valid sample of \experimentvalidsamplesize{} participants.
The summary of the demographics of our experiment sample ($N = \experimentvalidsamplesize{}$) is shown in the \autoref{sec: appendix-experiment-sample}.

\subsection{Methodological Limitations}

Our use of hypothetical scenarios as the background context may introduce some limitations.
Some participants might not have real-life experience with the specific tasks scenario. 
For example, a junior college student might not have worked in professional work context. 
And some participants might not use LLMs in the same scenario in real life.
To mitigate those problems raised from hypothetical scenario, we selected tasks that are generally common or easy to imagine.
Additionally, the participant's perceptions about the hypothetical scenarios may be sensitive to the wording of the description.
To mitigate the confounding impact on the judgement of different conditions, we have ensured that the descriptions are at similar length and levels of detail.
\revision{We also want to note that the hypothetical scenarios may not represent the way the task in the corresponding group is performed in real life, particularly for users who might choose not to use LLMs in the given scenarios.
We tried to mitigate the potential bias by comparing results across different groups all using the hypothetical scenarios.
However, the impact on the secret use rate of each scenario remains unknown.
Future studies could consider including users' willingness to use LLMs as a potential factor in their decision-making regarding secret use, and examine how users commonly perform these tasks in real life.}
Lastly, cognitive dissonance regarding LLM usage in the scenarios could also influence responses.
For example, if a participant's internal beliefs about the morality of using LLMs clash with their real-life behaviors, they might experience dissonance~\cite{festinger1962cognitive}.
To reduce dissonance, they might change their internal beliefs and claims or future actions~\cite{festinger1962cognitive}.
However, cognitive dissonance is a delicate psychological process and is not directly measurable.
We tried to mitigate its potential biases in our findings by focusing on cross-group comparisons.

\section{Study 2 Experiment: Results}

\subsection{Descriptive Statistics}

\subsubsection{Estimates of percentage of people who have intentions to keep the use of the LLMs secret}
\label{sec:desc-stats-secret-use-estimates}

For questions measuring the two forms of intentions to hide the use of LLMs (\autoref{sec:experiemnt-operationalization-dep-var}), we grouped the options ``Somewhat agree'', ``Agree'' or ``Strongly agree'' to estimate the percentage of people who expressed intentions to hide the use of LLMs. 
\autoref{tab:intentions-statistics} summarizes the results. 
Compared with the other five conditions, users in the reference condition ``General Information Search'' showed lower intentions on average in both passive non-disclosure (40\% v.s 72\%, \(SD = 0.12\)) and active concealment (16\% v.s 54\%, \(SD = 0.09\)).

\begin{table}[]
  \begin{threeparttable}
    \centering
    \caption{Estimates of the percentage of people who intend to keep the use of the LLMs secret (\%). 
    A participant is considered as likely to hide the use of LLMs if they chose ``Somewhat agree'', ``Agree'', or ``Strongly agree'' for the corresponding statement (presented in \autoref{sec:experiemnt-operationalization-dep-var}). 
    ``GIS''(General Information Search) is the reference condition, and the conditions that have significantly different adoption intentions in our linear regression analyses in \autoref{tab:linear-regression-results} are marked in bold.}
    \begin{tabular}{p{0.24\linewidth}p{0.08\linewidth}p{0.1\linewidth}p{0.1\linewidth}p{0.08\linewidth}p{0.08\linewidth}p{0.12\linewidth}}
    \toprule
    Dependent variable&GIS&Creative Writing& Academic Writing & School Works& Work Tasks&Social \newline Connection\\
    \midrule
    Passive non-disclosure&40.0&\textbf{66.0}**& \textbf{70.0}***& \textbf{62.0}**& \textbf{68.0}***&\textbf{92.0}***\\
    Active concealment&16.0&\textbf{52.0}***& \textbf{50.0}***& \textbf{50.0}***& \textbf{46.0}***&\textbf{70.0}***\\
 \bottomrule
    \end{tabular}
    \label{tab:intentions-statistics}
    \begin{tablenotes}
      \small
      \item **p<0.01; ***p<0.001.
    \end{tablenotes}
  \end{threeparttable}
\end{table}

\begin{figure}
    \centering
    \includegraphics[width=1\linewidth]{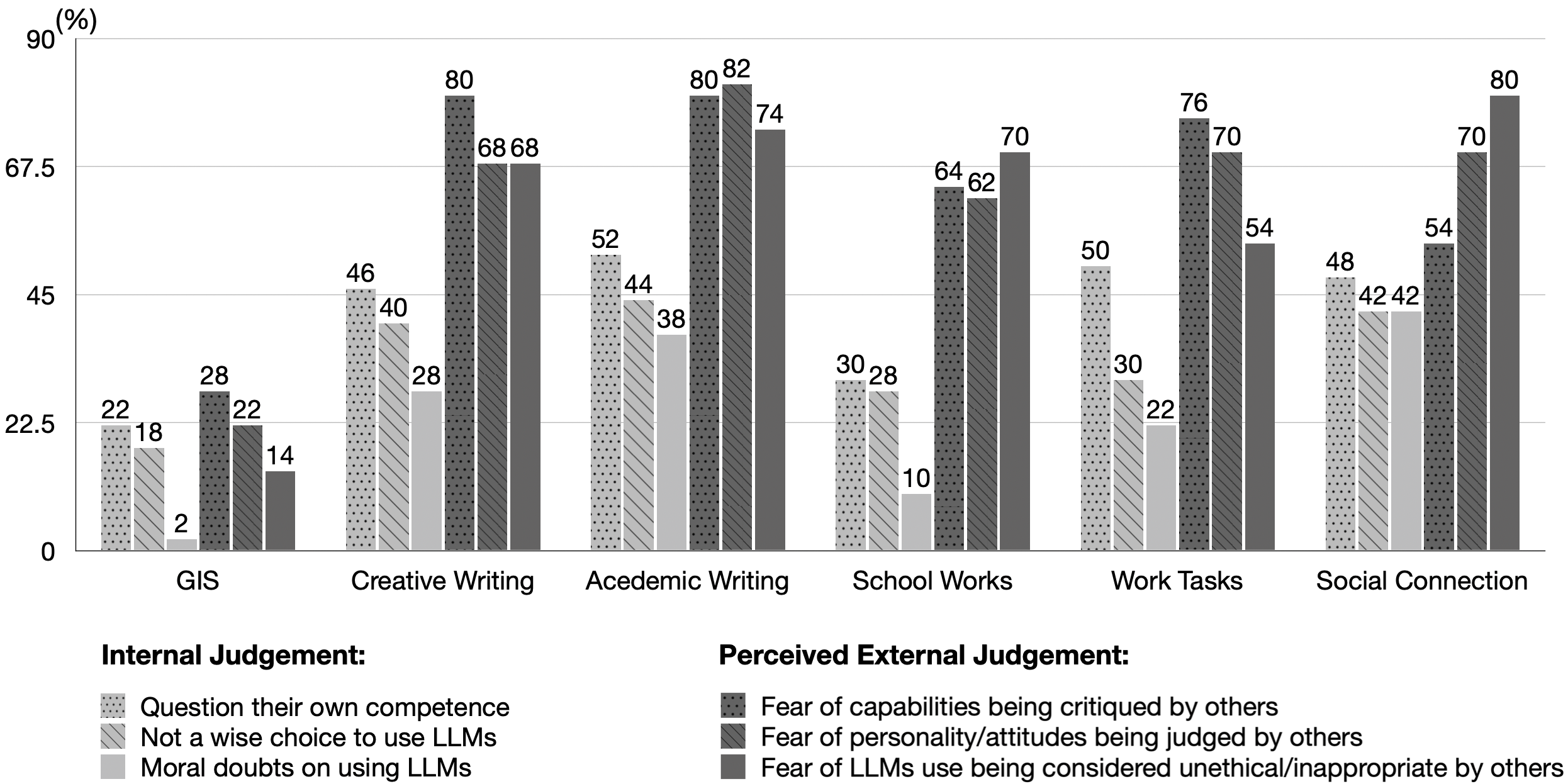}
    \caption{Estimates of the percentage of people who at least somewhat agree with the statements related to the internal judgement and the perceived external judgement on the use of LLMs. 
    ``GIS''(General Information Search) is the reference condition.}
    \label{fig:perceptions-statistics}
\end{figure}

\subsubsection{Estimates of users' internal judgement and perceived external judgement on the use of LLMs}
We also calculated the estimates of the two mediator variables across difference conditions. 
Results are shown in \autoref{fig:perceptions-statistics}.
In the ``General Information Search'' reference condition, agreement with both types of perceptions was much lower compared to the other five task conditions. 
This suggests that users exhibit more negative internal and external judgements towards using LLMs in the five specific tasks than in the reference condition. 
These findings are consistent with our survey results (\autoref{sec: survey-context-results}), where these five tasks frequently emerged in scenarios of secret use.
Notably, users exhibited more negative external judgements about LLM usage than internal judgements.
For instance, while only 10\% of participants at least somewhat agreed that using LLMs for school works raised moral doubts, 70\% believed that others would view this usage as unethical or inappropriate.

\subsection{Effects of Task Types and Individual Differences on the Intentions of Secret Use}
\label{sec: effects-task-individual-intentions}
We present the results for H1 and H2 based on our linear regression results.
We built two linear regression models for the two types of secret use, passive non-disclosure and active concealment, respectively (\autoref{tab:linear-regression-results}).
Multicollinearity was not a problem for all our linear regression analyzes because the maximum generalized variance inflation factor ($GVIF^{(1/(2*Df)}$) for our models is 1.41, which is lower than the cutoff value 2.25.

\begin{table}[]
\begin{threeparttable}
    \centering
    \caption{Linear regression results: The main effects of task types and individual differences on users' intentions of secret use of LLMs.}
    \begin{tabular}{p{0.4\linewidth} p{0.25\linewidth} p{0.2\linewidth}}
    \toprule
    \textbf{Independent variable}& Passive non-disclosure& Active concealment\\
 & Coef. (S.E.)&Coef. (S.E.)\\
    \midrule
 (Intercept)& 3.944*** (1.042)&4.071*** (1.117)\\
    \multicolumn{3}{l}{\textbf{Task types}}\\
        Task types (General info search=0)& & \\
        \hspace{3mm} Creative writing& 1.080** (0.354)& 1.393*** (0.380)\\
        \hspace{3mm} Academic writing& 1.300*** (0.355)& 1.652*** (0.381)\\
        \hspace{3mm} School work& 1.119** (0.355)&1.536*** (0.380)\\
        \hspace{3mm} Work tasks& 1.230*** (0.351)&1.374*** (0.377)\\
        \hspace{3mm} Social connection& 2.459*** (0.354)&2.444*** (0.380)\\
    \multicolumn{3}{l}{\textbf{Individual differences}}\\
        Self-esteem& -0.193 (0.206)& -0.168 (0.221)\\
        Self-efficacy& -0.036 (0.287)& -0.375 (0.308)\\Prosocialness& -0.138 (0.164)& -0.037 (0.176)\\General privacy concern& 0.222 (0.136)& 0.132 (0.146)\\
        Use frequency of LLMs& -0.033 (0.063)& -0.023 (0.067)\\
        Age& -0.008 (0.009)& -0.010 (0.010)\\
    \multicolumn{3}{l}{Gender (Male=0)}\\
        \hspace{3mm} Female& -0.068 (0.2132)& -0.158 (0.228)\\
        \hspace{3mm} Non-binary / third gender& -0.117 (0.596)&-0.308 (0.639)\\
        Education (Below bachelor=0) \\
        \hspace{3mm} Bachelor or Above & 0.157 (0.214)& 0.265 (0.230)\\
    \bottomrule
        \(R^2\)& 0.170&0.161\\
        Adjusted \(R^2\)& 0.129&0.120\\
    \bottomrule
    \end{tabular}
 \begin{tablenotes}
      \small
      \item  **p<0.01; ***p<0.001.
    \end{tablenotes}
  \label{tab:linear-regression-results}
  \end{threeparttable}  
\end{table}

We found that \textbf{the types of tasks involving LLMs significantly affect both forms of concealment intentions, supporting H1.}
More specifically, all five task conditions have positive significant effects on both intentions of passive non-disclosure and active concealment.
For example, the coefficient for \textit{Creative writing} condition for the intention of active concealment is 1.393, which represents an estimated increase in the 7-point active concealment intention rating if the task in which the LLMs used for is creative writing comparing with general information search.

In contrast, \textbf{individual differences did not show significant influences on the secret use intentions (H2.1-H2.5 are not supported).}
This finding suggests that tendencies towards secret use of LLMs are broadly consistent across various demographic and personality profiles.

\subsection{Mediation Effect of Internal Judgement and Perceived External Judgement}

\begin{figure}
    \centering
    \includegraphics[width=0.75\linewidth]{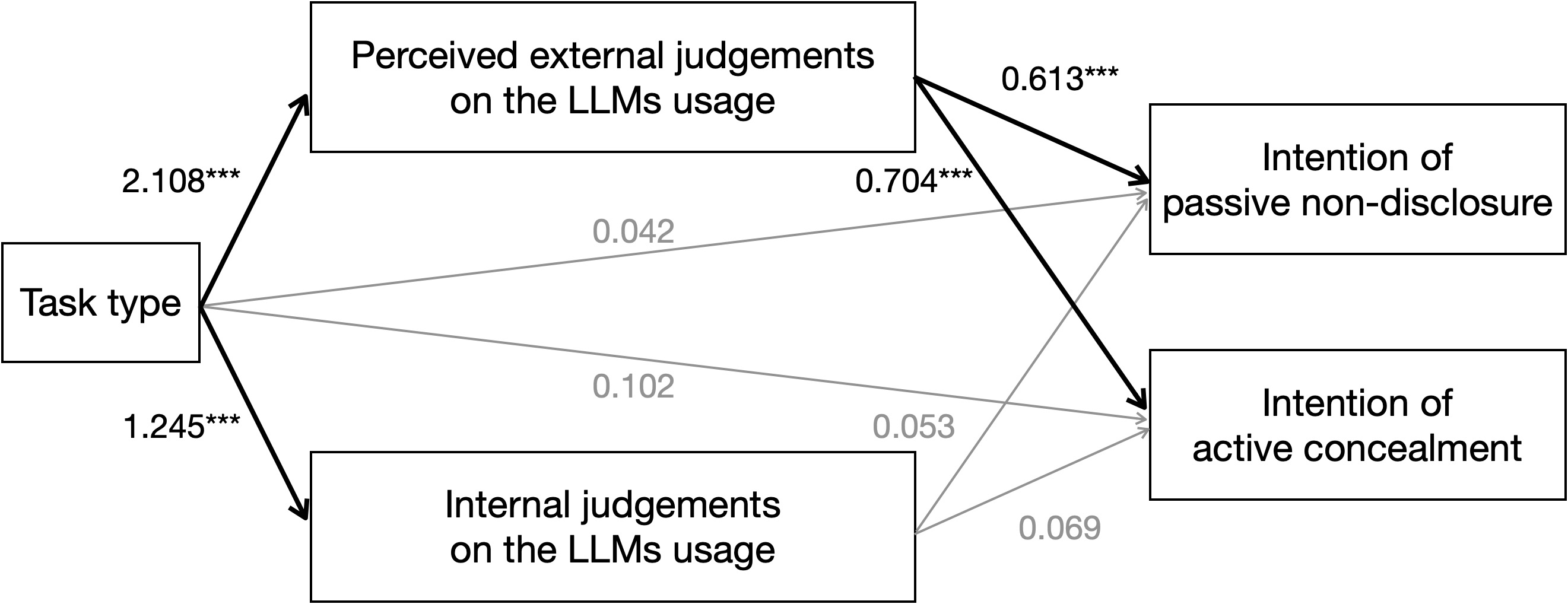}
    \caption{Structural equation model showing the mediation
analysis of task type, two types of perception on the use of LLMs and two forms of users' intentions to conceal the usage. Values represent standardized regression coefficients. It should be noted that both ``Perceived external judgement'' and ``Internal judgement'' in our study are all presented on negative side. Here, \revision{CFI: 0.935, SRMR: 0.042, RMSEA: 0.126}, p<0.001: ***}
    \label{fig:SEM-general-results}
\end{figure}

We conducted a mediation analysis using structural equation modeling (SEM) to test H3.1-3.2 about the mediation effect of uses' internal judgement and perceived external judgement.
In our mediation analysis, we focused on the relationship between the task type and the intention to keep the use of LLMs secret, which showed a significant effect in our linear regression model. 
\revision{In the SEM analysis, the task type was treated as a binary category with ``General Information Search'' set as 0 and other categories set as 1.
This approach allows us to determine whether task type could make significant differences compared with reference condition ``General Information Search'', which is essential to explore before breaking down the impact of different task types that could be further explored in the future.}
\revision{As shown in~\autoref{fig:SEM-general-results}, the Comparative Fit Index (CFI=0.935) and Standardized Root Mean Squared Residual (SRMR=0.042) indicate good model fit.
The high Root Mean Square Error of Approximation (RMSEA=0.126) indicates a poor fit.
Note that prior research suggests CFI is appropriate in more exploratory contexts, while RMSEA is appropriate in more confirmatory contexts~\cite{rigdon1996cfi}.
Our research takes the initial step in studying the secret use of AI, and hence is more exploratory in nature. However, the high RMSEA suggests complexities within the model that are worth further investigation in future research.
}

As shown in~\autoref{fig:SEM-general-results}, 
(1) Task type was significantly associated with both perceived external judgement \revision{(\(\beta\) = 2.108)} and internal judgement \revision{(\(\beta\) = 1.245)}. 
(2) Perceived external judgement on the LLM usage has significant impact on both intention of passive non-disclosure \revision{(\(\beta\) = 0.613)} and active concealment \revision{(\(\beta\) = 0.704)}. 
However, internal judgement showed non-significant effect on both two forms of intention.
(3) After controlling for perceived external judgements on the LLM usage, the direct effect of task type on the intention of passive non-disclosure \revision{(\(\beta\) = 0.042, \(p\) = 0.885)} and the intention of active concealment \revision{(\(\beta\) = 0.102, \(p\) = 0.737)} became non-significant.

\textbf{Hence, our results support H3.2, suggesting a significant mediation effect of perceived external judgment.}
The indirect effect of \textit{task types} through the \textit{perceived external judgement} accounts for \revision{88.8\%} of the total effect on the \textit{active concealment} intention, and \revision{92.3\%} on the \textit{passive non-disclosure} intention.
We did not find a significant mediation effect of internal judgment on the relationship between task type and intentions of secret use, so \textbf{H3.1 is not supported}.

\section{Discussion}

\subsection{The Secret Use of LLMs as a Prevalent Challenge for AI Transparency}

The findings of our study suggest that a large portion of the populations generally tend to hide the use of LLMs for various tasks (\autoref{sec:desc-stats-secret-use-estimates}).
This is a concerning issue, as we observe a significant overlap between the tasks in which users exhibited the secretive behaviors, and the tasks in critical domains such as academic writing, school work, creative writing, and professional work tasks.
As we have reviewed in \autoref{sec:community-level-requirements-of-transparency}, these domains have regulations or guidelines that explicitly require a disclosure of how LLMs are employed in achieving the tasks.
While having these regulations is the first step towards establishing clear rules and expectations for the transparency requirements of AI users, our study results suggest that non-compliance with these regulations may be prevalent.

When the use of LLM in critical tasks is not transparently disclosed, potential issues in the content generated by LLM, such as trade secret, bias or misinformation, can spread widely without sufficient precautionary measures being taken.
For example, using AI to research a court case has led to unwittingly including fictitious legal citations in a court brief~\cite{merken2023new}. 
In fact, our participants have shown some level of awareness of the potential negative consequences of their behavior. 
For example, EP80 (P80 in Study 2 - Experiment) thought~\blockquote{Using AI for detailed market research reports makes it feel both unreliable in a way.}, yet she still chose to keep her usage secret in both passive and active way. 
The seemingly contradictory behavior imply the existing challenges in ensuring that individual users do their part to provide the necessary transparency to prevent, mitigate, and diagnose the risks caused by AI-generated content.

Our findings regarding the mechanisms of users' intentions on the secret use of LLMs suggest that promoting transparency in LLM use cannot be a one-size-fits-all approach. 
Interventions must balance the need for AI transparency with individual privacy concerns, considering the context and varying levels of secretive behavior. 
Our findings emphasized the important role of users' perceived external judgment (perceived social norms) in shaping these behaviors and calls for tailored strategies to encourage responsible, transparent AI use by end-users.
\revision{Although we did not observe significant mediation effects from internal judgment as a whole in the quantitative analysis, qualitative findings from our first study did reveal some impact of users' internal judgment on their intentions to secret use, such as how using LLMs in a task affects their perceived self-efficacy and how it aligns with their own moral standards.
Our controlled experiment focused only on the mediation effect of two types of user perceptions towards LLM usage on their intention to secret use, yet there might be more complexities within the underlying mechanism.
Future research can look deeper into the mechanism of users' decision-making process in secret use based on our initial findings.
For example, including users' actual willingness to use LLMs in specific scenarios as a control variable can potentially provide a more comprehensive understanding of the decision-making process.}

\subsection{Addressing Tensions between Transparency Needs and Individual Privacy Concerns}

In our study, multiple tasks engendered high intentions for hiding the use of LLMs, while the potential harm levels associated with the lack of transparency varies.
Hence, we argue that it is beneficial to employ a multi-layered approach to encourage transparent disclosure of AI use among end-users.
In situations with less severe consequences, persuasive interventions like community policies and behaviorial nudges can guide or encourage transparent AI usage. 
For example, video platforms like YouTube and TikTok developed an AI labeling system to encourage creators to disclose their AI tool usage in video creation to audience~\cite{YouTube2024ResponsibleAI, TikTok2024AIGenerated}.
Our findings about the significant role of perceived external judgements suggests that social norm nudges~\cite{bicchieri2023norm} may be particularly relevant in addressing the users' concerns that lead to concealment intentions.
In addition, positive incentives such as a reward policy, could be more effective instead of only negative inhibition such as warning or punishment policy. 
This presents a promising direction for future research.
On the other hand, in critical areas where irresponsible usage could have severe consequences, or by adversarial users, soft nudges may not be enough. 
Therefore, in such high-stakes scenarios, stricter measures such as monitoring AI use, detecting AI-generated artifact, and implementing comprehensive policies and laws might be necessary.

\subsection{Emotional Stress Associated with the Secret Use of LLMs}
We want to note emotional stress as a potential negative impact of the secret use behavior on the user themselves.
Apart from contexts with more impact on the public domains, contexts like ``social connection'', where LLMs are used for personal messaging, caused the highest concealment intentions (92\% for passive non-disclosure; 70\% for active concealment). 
Qualitative responses from our survey and experiment suggest that users often experience emotional stress in these situations. 
For example, SP40 used ChatGPT for writing love letters to her spouse while keeping it secret. 
She said, ~\blockquote{...I felt guilty. My partner thought that I had said all of those kind words, but I did not}.
EP190 used an LLM tool to write a thank-you letter for a gift while she revealed bad feelings for herself,~\blockquote{I really did not feel involved enough to do so. It takes away from what it meant to write something to them if they were to know.}
In addition to the qualitative evidence, theoretical frameworks such as the self discrepancy theory also suggest that a conflict between users' self-beliefs about using LLMs for social messaging and their actual behavior may lead to self-discrepancies and subsequent emotional stress~\cite{carver1999self, scott1993self}.
Future research needs to be conducted to examine the issue of emotional stress in terms of how it may lead to changes in behaviors and long-term well-being.

\subsection{Towards the Establishment of Community Norms of AI Use and Disclosure}

Perceived external judgement on the use of LLMs, which was confirmed to be an important factor in shaping users' intentions of concealment of the usage, is a kind of perceived social norms towards this technology use~\cite{choi2013applying, higgs2015social}.
We are currently at the stage where LLMs emerge and open up new fields of human-AI interaction in various aspects of life.
Nevertheless, there remain uncertainties about how well users' perceptions align with actual social norms~\cite{diefenbach2019disrespectful,hechter2001have}.
If the actual reactions to LLM usage are positive or neutral, contrary to users' negative perceptions, it presents an opportunity to promote transparency by aligning users' perceived external judgments with the actual ones.
Therefore, we advocate going beyond just developing standards in a top-down manner, and emphasize the importance of engaging with a bottom-up approach to investigate people's perceptions of AI use in specific communities.
As an example, \citet{kapania2024m} conducted surveys and interview studies with HCI researchers to examine their perceptions of the ethical issues of LLM use in HCI research practices including ideation and project scoping, study design and execution, and analysis and paper writing.
These types of studies will play an instrumental role in understanding the community norms surrounding AI use.

\revision{Moreover, future research may need to look further into developing support for creating standardized, accurate, and meaningful disclosure of LLM use by systematically reviewing factors such as cultural differences, use domains, or varying levels of LLM operational involvement.
For example, the use of LLMs in personal vs. professional domains (e.g., maintaining personal relationships v.s. making professional medical decisions) or at different levels of involvement (e.g., refining language based on user drafts vs. generating the entire content) could potentially lead to different types or levels of risks.
The ideas of designing standardized, concise, and machine-readable privacy nutrition labels~\cite{kelley2009nutrition, kelley2010standardizing, emami2020ask}, as well as tools that make it easy to create these labels~\cite{li2022understanding, li2024matcha}, serve as a good example of addressing the transparency issues in data privacy. These concepts can be inspirational in addressing the transparency issues related to the secret use of LLMs.
Overall, we consider the design of support to facilitate community-level deliberation of norms and best practices of usage disclosure by AI users as a crucial and nascent area for future research.}

\section{Conclusion}

In this study, we explored the phenomenon of users' secret use of Large Language Models. 
We first delivered an exploratory survey among \surveyallusernumber{} users to qualitatively study real-world secret use cases on LLMs (n=\surveycasenumber{}) including the context and causes of such behavior.
Building on the survey findings, we conducted a controlled experiment among \experimentvalidsamplesize{} users to examine how the task type and individual differences influence users' intentions to conceal their LLM usage and the role of users' internal beliefs in mediating these effects.
Our results show that task types significantly impact users' intentions for both passive and active concealment, mostly through influencing users' perceived external judgement (e.g., perceived social norms) about using LLMs. 
Individual differences showed less impact on these secret use intentions.
Based on these findings, we discussed the challenges that secret LLM usage raises for AI transparency and proposed strategies that balance transparency needs with individual privacy concerns, advocating multi-layered and tailored intervention approaches to mitigate this issue. 
Emotional stress related to secretive AI usage emerged as a novel area for future research. 
Finally, we highlighted the need for conducting in-depth research at the community levels in establishing norms around LLM usage and disclosure, benefiting both individuals and the wider community.

\begin{acks}
This project is in part supported by National Science Foundation (NSF)
Grant IIS-2302730, National Institutes of Health (NIH) R01AI188576, and Northeastern University Tier-1 Research Grant. The content is solely the responsibility of the authors and does not necessarily represent the official views of the NSF, NIH, or Northeastern University.
\end{acks}

\bibliographystyle{ACM-Reference-Format}
\bibliography{sample-base.bib}

\appendix

\section{Demographics statistics of experiment sample (study 2)}
\label{sec: appendix-experiment-sample}
Demographics statistics of experiment sample can be reviewed in~\autoref{tab:demographics-statistics}.

\begin{table}[]
    \centering
    \caption{Demographics statistics of our experiment sample ($N = \experimentvalidsamplesize{}$).}
    \begin{tabular}{p{0.6\linewidth} p{0.1\linewidth} p{0.15\linewidth}}
    \toprule
    Demographic Characteristics& N& Sample (\%)\\
    \midrule
    \multicolumn{3}{l}{\textbf{Gender}}\\
        \hspace{3mm} Female& 143& 47.7\%\\
        \hspace{3mm} Male& 147& 49.0\%\\
        \hspace{3mm} Non-binary/third gender& 10& 3.3\%\\
    \multicolumn{3}{l}{\textbf{Age}}\\
        \hspace{3mm} 18-24& 53& 17.7\%\\
        \hspace{3mm} 25-34& 122& 40.7\%\\\hspace{3mm} 35-44& 66& 22.0\%\\\hspace{3mm} 45-54& 32& 10.7\%\\\hspace{3mm} 55-64& 23& 7.7\%\\\hspace{3mm} 65+& 4& 1.3\%\\
    \multicolumn{3}{l}{\textbf{Education}}\\
        \hspace{3mm} Bachelor’s degree or higher& 192& 64.0\%\\
    \multicolumn{3}{l}{\textbf{LLMs using frequency}}\\
        \hspace{3mm} Multiple times daily (several times throughout each day)& 19& 6.3\%\\
        \hspace{3mm} Daily (almost once every day)& 28& 9.3\%\\
        \hspace{3mm} Multiple times a week (but not daily)& 80& 26.7\%\\
        \hspace{3mm} Weekly (once a week)& 49& 16.3\%\\
        \hspace{3mm} Monthly (1-3 times a month)& 58& 19.3\%\\
        \hspace{3mm} Less than once a month& 28& 9.3\%\\
        \hspace{3mm} I've used it before, but not regularly& 38& 12.7\%\\
    \bottomrule
    \end{tabular}
    \label{tab:demographics-statistics}
\end{table}

\section{Task description overview (study 2)}
\label{sec: appendix-scenario-description}
Task descriptions can be reviewed in~\autoref{tab:experimental-manipulations}.

\begin{table}[]
    \centering
    \caption{Overview of task type manipulation in different hypothetical scenarios to establish distinct experimental conditions}
    \begin{tabular}{p{0.2\linewidth} p{0.75\linewidth}}
    \toprule
    Conditions& Descriptions shown to participants\\
    \midrule
    (Baseline) General information search& You knew the first Olympic Games were held in Greece. This sparked your curiosity: When did Greece host the Olympics again after the first one? So, you searched this information through an AI chatbot.\\
    Creative writing& You’re an amateur writer who publishes novels on popular online platforms to reach a broader audience. Aware of AI chatbot’s advanced writing capacities, you used it for your novel writing, including brainstorming fresh ideas, structuring your plot, and refining the language to make your stories stand out.\\
    Academic writing& You’re a college student on a research track. You needed to have a paper ready for publication submission by the end of the semester. Aware of AI chatbot’s impressive abilities in academic writing, you used it to help your paper, including brainstorming the topic, drafting some parts, and refining the language.\\
    School work& You’re a college student with a crucial assignment due in just a few days, alongside other coursework. During this crunch time, you found the AI chatbot could be a lifesaver. So you used it to help organize your ideas, draft your arguments, and refine the language.\\
    Work tasks& You’re a marketing analyst. As part of your job responsibilities, you have recently been assigned several tasks involving the preparation of detailed market research reports. Aware of AI chatbot's powerful abilities in synthesizing information and generating clear, insightful content, you decided to use it to assist with this work task.\\
    Social connection& You’ve received a wedding invitation from your friend for next weekend. Right now, you’re super busy with work, and you found yourself overwhelmed to find the time to write a thoughtful reply. So you listed key points and asked the AI chatbot to generate a reply email for your friend.\\
    \bottomrule
    \end{tabular}
    \label{tab:experimental-manipulations}
\end{table}

\end{document}